\def\year{2019}\relax
\documentclass[letterpaper]{article} 
\usepackage{aaai19}  
\usepackage{times}  
\usepackage{helvet}  
\usepackage{courier}  
\usepackage{url}  
\usepackage{graphicx}  
\usepackage{amsmath}
\usepackage{algorithmic}
\usepackage{multirow}
\usepackage{algorithm}
\usepackage{array}

\newlength\savedwidth
\newcommand\whline{\noalign{\global\savedwidth\arrayrulewidth
                            \global\arrayrulewidth 2pt}%
                   \hline
                   \noalign{\global\arrayrulewidth\savedwidth}}

\begin{document}
%
\title{TableSense:  Spreadsheet Table Detection with Convolutional Neural Networks}

\author{Haoyu Dong\textsuperscript{1},
Shijie Liu\textsuperscript{2},
Shi Han\textsuperscript{1},
Zhouyu Fu\textsuperscript{1},
Dongmei Zhang\textsuperscript{1}\\
\textsuperscript{1}{Microsoft Research, Beijing 100080, China.}\\
\textsuperscript{2}{Beihang University, Beijing 100191, China}\\
\{hadong, shihan, zhofu, dongmeiz\}@microsoft.com,
shijie\_liu@buaa.edu.cn}

\maketitle

\begin{abstract}
Spreadsheet table detection is the task of detecting all tables on a given sheet and locating their respective ranges. 
Automatic table detection is a key enabling technique and an initial step in spreadsheet data intelligence. 
However, the detection task is challenged by the diversity of table structures and table layouts on the spreadsheet. 
Considering the analogy between a cell matrix as spreadsheet and a pixel matrix as image, and encouraged by the successful application of Convolutional Neural Networks (CNN) in computer vision, we have developed TableSense, a novel end-to-end framework for spreadsheet table detection. 
First, we devise an effective cell featurization scheme to better leverage the rich information in each cell; 
second, we develop an enhanced convolutional neural network model for table detection to meet the domain-specific requirement on precise table boundary detection; 
third, we propose an effective uncertainty metric to guide an active learning based smart sampling algorithm, which enables the efficient build-up of a training dataset with 22,176 tables on 10,220 sheets with broad coverage of diverse table structures and layouts. 
Our evaluation shows that TableSense is highly effective with 91.3\% recall and 86.5\% precision in EoB-2 metric, a significant improvement over both the current detection algorithm that are used in commodity spreadsheet tools and state-of-the-art convolutional neural networks in computer vision.
\end{abstract}

\section{Introduction}
Spreadsheets are a critical end-user development tool for data management and analysis. 
In spreadsheet data, the table is a key structure for data processing and information presentation. 
Automatic table detection is an important initial step for one-click intelligence features such as Ideas in Excel or Explore in Google Sheets, where insights can be recommended from the detected tables with an automated end-to-end experience. It is also a key enabling technique for spreadsheet intelligence on the cloud or mobile devices. On the cloud, a vast variety of data are available for batch processing while there is no user interface to specify table ranges. On mobile devices, user navigations and interactions for specifying table ranges are restricted to a small screen.

\begin{figure*}[!t]
\centering
\includegraphics[width=5.1in]{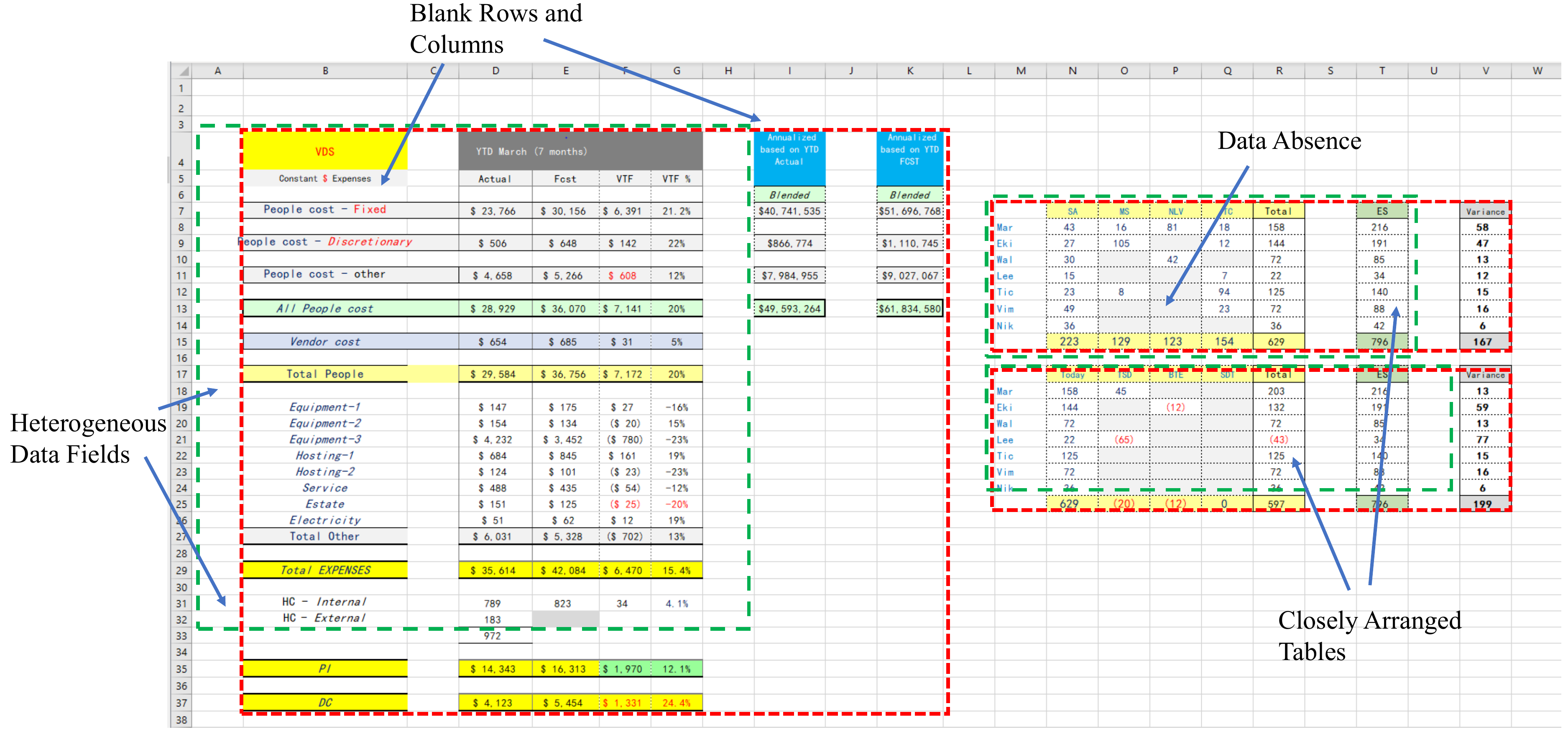}
\caption{A sample spreadsheet with three tables showing various artifacts.
 Dotted red bounding boxes and dashed green bounding boxes show the tables detected by
 TableSense and Mask R-CNN, respectively.}
\label{fig1}
\end{figure*}

Despite the importance of automatic table detection for spreadsheets, this problem has largely been overlooked for decades in both the research community and industry. 
Previous research on table detection has mainly targeted other medias, e.g. HTML \cite{wang2012understanding,zhai2005web,wang2002machine}, images \cite{gatos2005automatic,zuyev1997table,hu1999medium,liu2008identifying,shafait2010table} and PDFs \cite{fang2012table,liu2007tableseer}. 
The aim is to retrieve (mostly likely single) table regions from ambient text. 
The major challenge for these techniques is the understanding of binary files based on meta data analysis and image processing, but the table boundaries are clear. 
The scenario with spreadsheet table detection is fundamentally different. 
As demonstrated in Figure \ref{fig1}, a single sheet can have multiple tables cluttered around with potentially different structures for each table. 
The diversity in multi-table layout and structure significantly confounds the problem with obfuscated table boundaries. 
To the best of our knowledge, there is no prior research effort on this problem in academia, while region-growth techniques are commonly used in commodity spreadsheet tools. However, region-growth is quite fragile with the presence of complicated table structures and layouts on the sheet. For example, it fails on the sample sheet shown in Figure \ref{fig1}.

A key insight from inspecting our spreadsheet corpus is that the vast majority of tables are ``human-friendly" but not ``machine-friendly". Only less than 3\% of the tables in our corpus have a pre-defined data model or are properly normalized for automated analysis. Most tables are designed and crafted for summarization or reporting purposes, with customized table designs by end users to fit their specific goals. This often leads to considerable diversity in such tables and their layouts on a sheet, with artifacts to improve clarity and visual quality. Some of these artifacts are shown in Figure\ref{fig1}. For the table at range B4:K37\footnote{By convention, a contiguous cell range is identified by addresses of the cells in the top-left and bottom-right corners separated by `:'. Cells are referenced by the column and row indices.}, range B31:G33 is embedded in the larger homogeneous range of the table, aligning to the same shared header but cutting the homogeneous range into two sections. It is not uncommon for heterogeneous data fields to be inserted into a homogeneous range intentionally to improve summary. It is also typical for users to intentionally insert blank rows or columns inside a table to improve alignment and clarity. Furthermore, users tend to arrange related tables with similar structures closely with a small gap between them, which is the case for the second table at range M7:V15 and the third table at range M17:V25 in Figure \ref{fig1}. While these designs make the spreadsheet tables more ``human-friendly'', they also make the table regions less coherent and table boundaries more ambiguous. This significantly challenges spreadsheet table detection, making it difficult to achieve desirable accuracy based on simple region-growth or conventional machine learning methods.

Alternatively, a sheet can be viewed as a two-dimensional array of cells, and a table is a subset of cells occupying a contiguous range on the sheet. This motivates a distinct approach by leveraging convolutional neural networks \cite{girshick2014rich,uijlings2013selective,hosang2016makes,lecun1989backpropagation,krizhevsky2012imagenet,girshick2015fast,ren2015faster} to capture spatial correlations and learn high-level representations of cell matrices from vast variety of real-world spreadsheets. Nevertheless, the cross-domain application of CNNs is never straightforward due to domain-specific characteristics of spreadsheet data and tasks. Directly applying CNNs to spreadsheet data without incorporating domain-specific and task-specific cues fails to achieve desirable accuracy. 
In this paper, we propose TableSense, a CNN model with several key enhancements customized for spreadsheet table detection. In TableSense, there are three major technical contributions to address the following domain-specific challenges.

1.	Unlike pixels in images, there is no canonical representation for cells in spreadsheets. Cells typically contain much richer information such as data types, data formats, cell formats, formulas, etc. We have designed an effective featurization scheme to capture cell information to initiate end-to-end model learning.

2.	Unlike object detection in images, table detection in spreadsheets requires precise boundary segmentation. Leaving one or two lines of pixels out of a region does not invalidate a correct object detection in an image, but leaving a single line of cells out of a table region leads to incorrect table detection results on a sheet. In TableSense, we proposed a key enhancement for refining the table boundaries.

3.	Unlike image data, which is easy to collect and to label, spreadsheet data is of much smaller scale and is thus much sparser regarding to the variety of table structures. Labeling the tables for spreadsheet is also more difficult and time-consuming as the human labeler often needs to understand the meaning of the table for correct labeling. Therefore, training set selection for labeling requires smart sampling from the limited dataset to quickly achieve desirable coverage of various table structures and optimize sample efficiency for training. In our method, we adopt the active learning approach to label data in iterations, where we define an effective uncertainty metric as key to selecting the least confident sheets to label in the next iteration.

We have built a working system for spreadsheet table detection based on the TableSense technology. Our evaluation shows that TableSense is highly effective. It achieves 91.3\% recall and 86.5\% precision measured by our proposed EoB-2 metric, a significant improvement over region-growth algorithms that are used in commodity spreadsheet tools (58.5\% recall and 55.2\% precision) and the state-of-the-art convolutional neural networks used in computer vision tasks.

\section{Preliminaries}
\subsection{Problem Statement}
Spreadsheet table detection is the task of detecting all tables on a given sheet and locating their respective ranges. In such a task, an input sheet is represented by a matrix of cells. The output is a list of tables detected, where the range of each detected table is represented by a 4-tuple $(\text{col}_\text{left},\text{row}_\text{top},\text{col}_\text{right},\text{row}_\text{bottom})$, which specifies the $x$ and $y$ coordinates for the top-left and bottom-right corners of the bounding box (bbox).

\subsection{Performance Metric}
For the object detection task in images, a successful detection is usually measured by the Intersection-over-Union (a.k.a. IoU) metric. Intuitively, given a detected bounding box B and its corresponding ground truth bounding box $B^{'}$, IoU measures the area of intersection against the area of their union, i.e., 

\vspace{-0.2cm}

\begin{equation}
\text{IoU} = \frac{\text{area}(B\bigcap B^{'})}{\text{area}(B \bigcup B^{'})}
\end{equation}

A threshold of $\text{IoU}>0.5$ has been commonly used to indicate successful detection in object detection.
However, for spreadsheet table detection, the IoU threshold of 0.5 is too loose to be useful. Since table detection is a preliminary step for intelligent spreadsheet data analysis, subsequent steps would be applied to the detection results to perform table structure analysis and data summaries discovery\footnote{https://support.office.com/en-us/article/ideas-in-excel-3223aab8-f543-4fda-85ed-76bb0295ffc4}. Hence, in general, the task requires more precise bounding box detection for a table.

Therefore, we define an Error-of-Boundary (EoB) metric to measure how precisely the detection result is aligned to the ground truth bounding box.

\vspace{-0.4cm}

\begin{equation}
\begin{aligned}
\text{EoB}  = & \max(|\text{row}_\text{top}^B-\text{row}_\text{top}^{B^{'}}|, 
|\text{row}_\text{bottom}^B-\text{row}_\text{bottom}^{B^{'}}|, \\
& |\text{col}_\text{left}^B-\text{col}_\text{left}^{B^{'}}|,
 |\text{col}_\text{right}^B-\text{col}_\text{right}^{B^{'}}|)
\end{aligned}
\end{equation}

\text{EoB}$==$0 means the detected bounding box exactly matches the ground truth, while \text{EoB}$\le$2 is good enough to flag the successful detection of a table on a sheet. The tolerance of 2 is chosen by accounting for the existence of titles, footnotes and side notes around the exact table region, and one can employ existing techniques such as \cite{krishnakumar2007active} to distinguish them given our detection results within the specified tolerance. In this paper, we report precision and recall with both EoB$\le$2 (EoB-2) and EoB$==$0 (EoB-0).

\subsection{Datasets}

All experimental data in the development of TableSense is from our \textbf{WebSheet} dataset, which is a web-crawled spreadsheet corpus including 4,290,022 sheets.

\textbf{WebSheet10k} is a sampled subset of \textbf{WebSheet} for human labeling. It contains 10,220 sheets in English, where all table regions on each sheet have been labeled with a corresponding bounding box. To control labeling quality, each sheet has been labeled by a human labeler and then verified by another human labeler. To ensure high coverage of various table structures and multi-table layouts on sheets, we adopt an active learning framework 
to build \textbf{WebSheet10k} in iterations. Details are provided in Section \ref{activelearning}. 

\textbf{WebSheet400} is our test set with labels, which contains 400 randomly sampled sheets with 795 tables from WebSheet without any overlap with \textbf{WebSheet10k}. 

Note that, solving release and compliance issues for all files in WebSheet  needs considerable efforts, so we recently annotate and publish table regions for VEnron2, VEUSUS, and VFUSE to facilitate recent research\footnote{https://github.com/microsoft/TableSense}.

\begin{figure*}[t]
\centering
\includegraphics[width=6.5in]{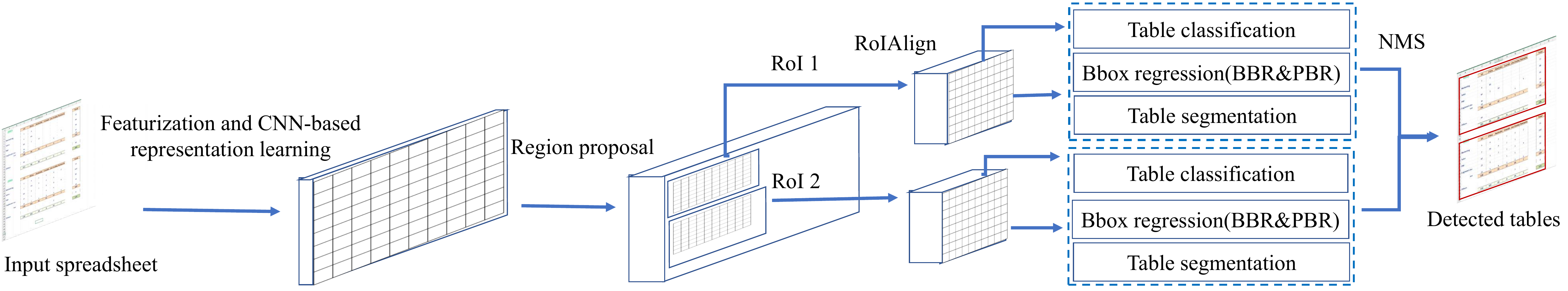}
\caption{Framework of TableSense for spreadsheet table detection.} \label{fig2}
\end{figure*}

\section{TableSense Suite}

Most spreadsheet documents are created by end users for human inspection. To enhance human readability, various visual attributes, such as colors and border lines, and semantic attributes, such as text formats and formulas, are usually used to create tables. The great diversity in human tabulation styles poses great challenges for approaches based on heuristics or shallow learning models to fully capture the intrinsic table features. Alternatively, a sheet can be viewed as a two-dimensional array of cells, and a table is a subset of cells occupying a contiguous range on the sheet. This motivates a distinct approach by leveraging convolutional neural networks to capture spatial correlations and learn high-level representations of sheet cells.

Nevertheless, the table detection task has some domain specific characteristics. As demonstrated in Evaluation Results, direct application of standard object detection methods fails to achieve satisfactory accuracy. Hence, we customize the CNN framework for table detection, and discuss three key approaches for the successful integration of deep CNN into our detection framework, including cell featurization, model enhancement with precise bounding box regression and active learning for efficient data labeling.

\subsection{TableSense Framework} \label{secmask}

Figure \ref{fig2} presents our framework tailored for table detection.  It is an end-to-end model containing a series of modules as follows.

\begin{itemize}
\item  \textbf{Cell featurization}: Since cells do not have a canonical representation in the spreadsheet, we need to extract cell features before feeding them to the pipeline. Details of cell featurization will be provided in 
	Section \ref{cellfeat}.
\item  \textbf{CNN backbone}: CNN is the backbone of our framework to capture spatial correlations and learn high-level representations from input cell matrix \cite{girshick2014rich,uijlings2013selective}, and fully convolutional network is adopted here so as to enable the model to process spreadsheets of various sizes without rescaling them.
\item  \textbf{Table detection head}: The two-stage detection mechanism which achieves state-of-the-art results in computer vision \cite{ren2015faster,girshick2014rich,he2017mask} is adopted. In this module, the feature maps generated by the CNN backbone are fed to a Region Proposal Network (RPN), which further produces a list of Regions of Interest (RoIs). Then RoIAlign \cite{he2017mask} extracts feature maps from each RoI for bounding box regression. Then a CNN-based bounding box regression branch \cite{girshick2015fast} refines the boundaries of these RoIs, a CNN-based table classifier simultaneously scores these RoIs, and a segmentation branch generates the cell-level table mask. These branches are applied to each RoI separately. Finally, Non-Maximum Suppression (NMS) is used to rank the bounding boxes and filter redundant ones \cite{ren2015faster,he2017mask}. For our task, RoIAlign which is based on bilinear interpolation can preserve more precise per-cell correspondence than RoIPool \cite{ren2015faster} which uses simple hard quantization. 
\end{itemize}

In this paper, we enhance the Bounding Box Regression (BBR) with a novel Precise Bounding box Regression (PBR) to achieve precise table boundaries.

\subsection{Cell Featurization} \label{cellfeat}

\begin{table}
\centering
\caption{20 features employed for cell featurization.}\label{tab2}
\scalebox{0.85}{
\begin{tabular}{l l}
\whline

\textbf{Description} & \textbf{Feature value} \\
\hline
\multicolumn{2}{l}{ \textbf{Value string}} \\

	If the string is non-empty.  & \{0, 1\} \\
 Length of the string. & Integer\\
 Proportion of digits in the string. & [0.0, 1.0] \\
Proportion of letters in the string & [0.0, 1.0] \\
 If percent symbol (``\%") exists in the string.  &   {0, 1}\\
 If decimal point (``.") exists in the value  &   {0, 1}\\
\hline
\multicolumn{2}{l}{ \textbf{Data format}} \\

If data format matches a numerical template &   {0, 1}\\
If data format matches a date template &   {0, 1}\\
If data format matches a time template &   {0, 1}\\
Length of the matched template string, if any.  &   Integer\\
\hline
\multicolumn{2}{l}{ \textbf{Cell format}} \\

 Background fill color &   Categorical\\
 Font color& Categorical\\
  If bold font is applied.  &   {0, 1}\\
 If the cell has left border.  &   {0, 1}\\
 If the cell has top border.  &   {0, 1}\\
If the cell has right border.  &   {0, 1}\\
If the cell has bottom border. &   {0, 1}\\
If the cell is merged with horizontal neighbor.  &   {0, 1}\\
 If the cell is merged with vertical neighbor.  &   {0, 1}\\
\hline
\multicolumn{2}{l} {\textbf{Formula}} \\

 If formula exists in the cell.  &   {0, 1}\\
\whline

\end{tabular}}
\end{table}

Cells in a spreadsheet correspond to pixels in an image, but they encode much richer information than pixels do. Whether such information is well extracted and leveraged can lead to remarkable differences in the accuracy of table detection, as reported in Evaluation Results. Therefore, cell featurization is an additional but important initial step in TableSense.
In general, there are four major information sources of a cell, i.e., value string, data format, cell format, and formula. While value string and cell format are visually perceivable to users, data format and formula are latent unless users explicitly explore them. We have identified 20 features as shown in Table \ref{tab2} from the above information sources. Each feature acts as a separate channel in the input layer. If the input spreadsheet is a matrix of $h\times w$ cells, then the input will be a $h\times w\times 20$ tensor, to which the convolution operations are directly applied.

Our insights behind the feature set are briefly summarized as follows. Unlike free-form textures in an image object, a spreadsheet table has its unique characteristic of being essentially a composition of vertically/horizontally expanded components, which are headers, data fields and data records. Identifying cell-level cohesion along the expanding direction of such components and detecting cell-level contrast across components would intrinsically help identify those components and further scope the table range. Consistency in data formats, formulas, and other statistics are all effective information for the detection of such cell-level cohesion and contrast.

\begin{figure}[t]
\centering
\includegraphics[width=3.3in]{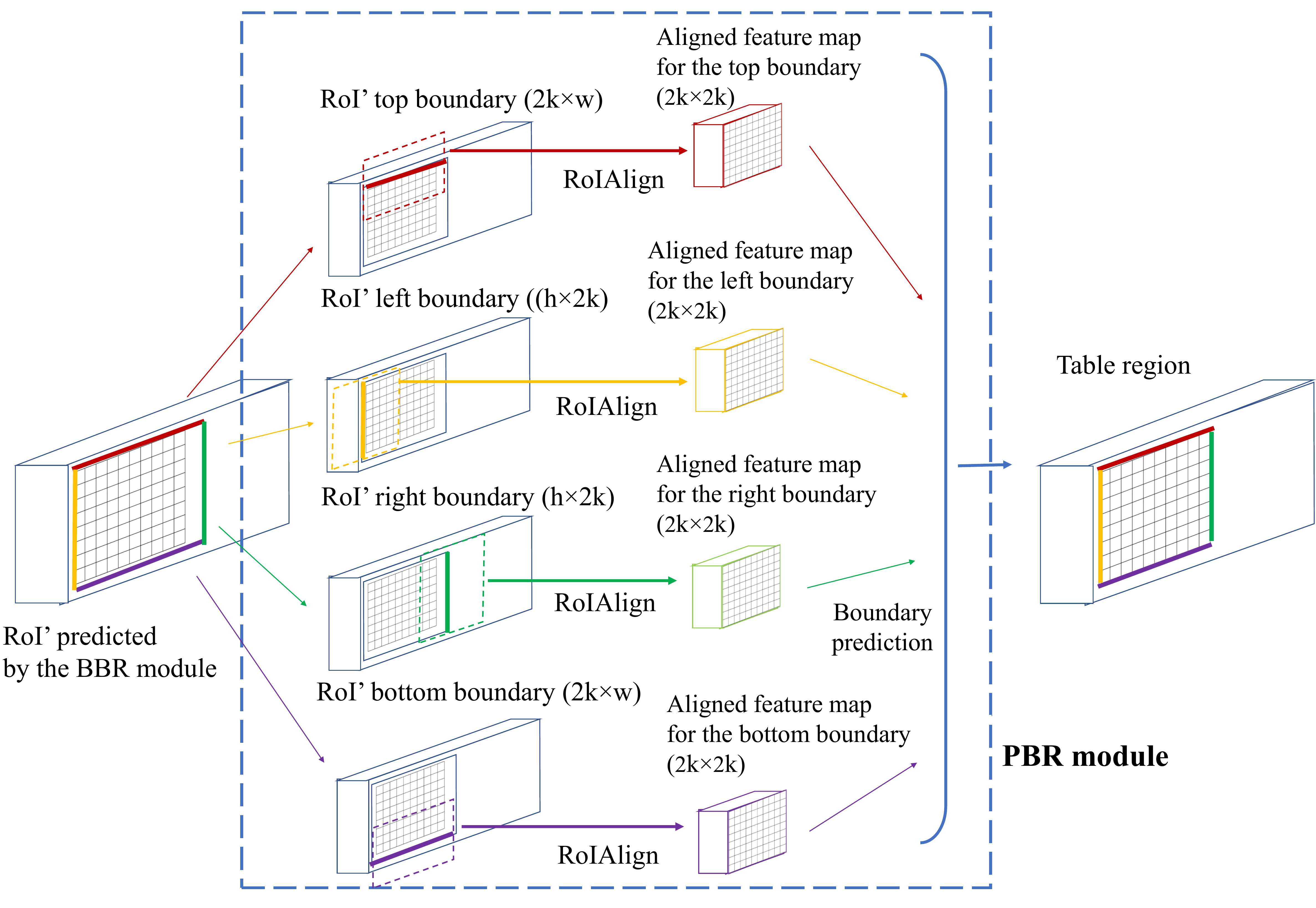}
\caption{Framework of PBR module in TableSense.} \label{fig3}
\end{figure}

\subsection{Precise Bounding Box Regression for TableSense}
The Bounding Box Regression (BBR) module in Faster R-CNN \cite{girshick2015fast} is optimized for object detection. In object detection, we do not care much about the precise bounding box locations, but rather the overlapping ratio between the detected bounding box and ground truth. This is modeled by the BBR cost function below:  

\begin{equation}
L_\text{reg}(t,t^*)=\sum_{i \in \{x,y,w,h\}} \text{smooth}_{\text{L}_1}(t_i - t_i^*)
\end{equation}

\begin{equation}
\text{smooth}_{\text{L}_1}(x)=
\left\{
\begin{array}{lr}
0.5x^2 \text{,      } & \text{if} |x|<1, \\
|x|-0.5  \text{,}& \text{otherwise}, 
\end{array}
\right.
\end{equation}

where $\text{smooth}_{\text{L}_1}()$ is the smooth $\text{L}_1$ loss function defined in \cite{girshick2015fast}, $t_x$, $t_w$ , ${t_x}^*$ and ${t_w}^*$ are relative coordinates with the encoding scheme below (likewise for $t_y$, $t_h$, ${t}^*{_y}$ and ${t}^*{_h}$):

\begin{equation}
\begin{aligned}       
& t_x=(x-x_a)/w_a,        \text{                   } t_w=\log(w/w_a) \\
& {t_x}^*=(x^*-x_a)/w_a,  \text{            } {t_w}^*=\log(w^*/w_a)\\
\end{aligned}
\end{equation}

where $x$, $y$, $h$ and $w$ denote the centroid coordinates, width and height of the predicted bounding box, $x_a$ and $x^*$ denote the x-coordinates for the anchor box and ground truth box respectively (likewise for $y$, $w$, $h$). This can be thought of as a bounding-box regression from an anchor box to a nearby ground-truth box. 

The cost function defined above is designed to minimize the relative error instead of absolute deviations. This leads to distinct variable updating behaviors for different sized anchor boxes. We show this for $x$ and $w$ below, and the same analysis also holds for $y$ and $h$. The gradient of $L_\text{reg}$ at $x$ and $w$ is as follows: 

\begin{equation}
\frac{\partial{L_\text{reg}(t,t^*)}}{\partial{x}}=
\left\{
\begin{array}{lr}
(x-x^*)/{w_a}^2, & |x-x^*|<w_a, \\
1/w_a, & x-x^* \ge w_a, \\
-1/w_a, & x-x^*\le -w_a, \\
\end{array}
\right.
\end{equation}

\begin{equation}
\frac{\partial{L_\text{reg}(t,t^*)}}{\partial{w}}=
\left\{
\begin{array}{lr}
\ln \frac{w}{w^*}/w  ,   & \frac{w^*}{e}<w<ew^*, \\
1/w, &  w \ge ew^*, \\
-1/w, & w \le \frac{w^*}{e}, \\
\end{array}
\right.
\end{equation}

Note the gradient at $x$ is at least inversely proportional to $w_a$, and the gradient at $w$ is inversely proportional to $w$, since $\ln(w/w*)$ is small. Hence, the larger anchor box width  $w_a$, the smaller step variable $x$ is updated in the backward pass, even though the absolute deviation $x-x^*$ may still be large. Likewise, $w$ also takes a small update step if $w$ is large, even though $w$ may still be far from $w^*$. Hence, the BBR loss lacks sensitivity for absolute errors in large tables compared with relative errors in small tables. In addition, the downsampling strategy in standard RoIPool and RoIAlign module also leads to potential accuracy loss.

To address the above problem, we propose a new Precise Bounding Box Regression (PBR) module to extend the standard BBR in Faster R-CNN. As shown in Figure \ref{fig3}, the PBR module implements a cell-level boundary regression for each direction separately based on the corresponding local receptive field with a size of $h$$\times$$2k$ for horizontal direction and $2k$$\times$$w$ for vertical direction, where $k$ controls the maximum tolerance on absolute deviation for PBR regression. This mechanism prevents the RoIAlign module from downsampling cell features in the regression direction, and effectively preserves cell-level precision for boundary regression. We combine BBR with PBR to predict the precise table boundaries in a coarse-to-fine fashion, where BBR is applied to probe for the approximate table boundaries, and PBR is employed to refine the result returned by BBR. In contrast, it would be difficult to find the exact boundaries using standard BBR in isolation as discussed above. On the other hand, using PBR alone may have convergence issues when there is a large deviation between the initial location and ground truth which we will discuss shortly. 

We define new regression targets $t_{\text{left}}$, $t_{\text{right}}$ and ground truth ${{t^*_\text{left}}}$, ${{t^*_\text{right}}}$ for PBR in terms of absolute deviations from the anchor (likewise for $t_\text{top}$, $t_\text{bottom}$, ${{t^*_\text{top}}}$ and ${t^*_\text{bottom}}$). 

\begin{equation}
\begin{aligned}       
& t_\text{left}=x-x_a-w/2, \;\;t_\text{right}=x-x_a+w/2  \\
& {t^{*}_\text{left}}=x^*-x_a-w^*/2, \;\;{t^*_\text{right}}=x^*-x_a+w^*/2
\end{aligned}
\end{equation} 
                                  
The purpose of PBR is to minimize the absolute deviations between predicted boundary boxes and their ground truth values. It can be formulated as the cost function below: 

\begin{equation}
L_\text{PBR}(t,t^*)=\sum_{i \in \{\text{top},\text{bottom},\text{left},\text{right}\}} R(t_i-{t^*_i})
\end{equation}

\begin{equation}
R(x)=
\left\{
\begin{array}{lr}
0.5x^2 \text{,      } & \text{if } |x|<k, \\
0.5k^2  \text{,}& \text{otherwise}, 
\end{array}
\right.
\end{equation}
where $R(x)$ denotes the robust loss function for absolute deviations, and parameter $k$ controls the maximum tolerance on absolute deviation for PBR regression. The loss increases monotonically with deviations less than $k$ columns/rows and the same loss is incurred for any deviation over $k$. The loss $R(x)$ is well suited for precise boundary detection, since any mismatch between detected boundaries and the ground truth is undesirable, but larger deviations should not be over-penalized either. 

In this task, too small $k$ would limit PBR's receptive field and cause failure. But for large $k$, if the receptive field of a single PBR contains multiple top (similar for left, right, bottom) boundaries from multiple enclosed table regions, the PBR layer may fit to the mismatched one at test time. To reduce such risks, a moderate value of $k$ is preferred. Furthermore, because of the limitation of $k$, the initial bounding box should be reasonably close to the ground truth for $R(x)$ to be effective. Hence BBR is adopted for coarse localization before the use of PBR, and it also makes PBR converge faster. In theory, we could cascade multiple PBRs before BBR for a larger receptive field.

With the proposed PBR module, we can see that the variable updating behaviors are dependent on the absolute deviation independent of the size of the bounding box. We show this for $x$ and $w$ below, and the same analysis also holds for $y$ and $h$. The gradients of $L_\text{PBR}$ on $x$ and $w$ are as follows.

\begin{equation}
\begin{aligned}
& \frac{\partial{L_\text{PBR}(t,t^*)}}{\partial{x}}= \left( {t_\text{left}} - {t^{*}_\text{left}}  \right) \text{sign}\left( k-| {t_\text{left}} - {t^{*}_\text{left}}| \right) \\
& +  \left( {t_\text{right}} - {t^{*}_\text{right}}  \right) \text{sign}\left( k-| {t_\text{right}} - {t^{*}_\text{right}}| \right)
\end{aligned}
\end{equation}

\begin{equation}
\begin{aligned}
& \frac{\partial{L_\text{PBR}(t,t^*)}}{\partial{w}}= \frac{( {t_\text{left}} - {t^{*}_\text{left}}  )}{2} \text{sign}\left( k-| {t_\text{left}} - {t^{*}_\text{left}}| \right) \\
& +  \frac{( {t_\text{right}} - {t^{*}_\text{right}}  )}{2} \text{sign}\left( k-| {t_\text{right}} - {t^{*}_\text{right}}| \right)
\end{aligned}
\end{equation}

The gradients above indicate that the updates for $x$ and $w$ only depend on the absolute deviations $x-x^*$ and $w-w^*$. Since the prediction targets are not normalized by $w$ or rescaled by logarithm, the PBR module is better suited for precise boundary prediction. 

The losses of BBR and PBR are added together for end-to-end training. While BBR and PBR both employ CNNs as their backbones, different prediction targets, loss functions, receptive fields and RoIAlign targets are adopted. The PBR module can effectively refine the BBR predictions and further significantly improves localization accuracy.

\subsection{Active Learning Framework} \label{activelearning}

Training a table detection system requires a large amount of labeled spreadsheet data. Since it is unrealistic to label all sheets due to the amount of time and labor cost involved, we adopt an active learning framework to label sheets in iterations, thus minimizing the labeling cost while maximizing learning performance. The main assumption behind active learning is that if an active learner can freely select any samples it wants, it can outperform random sampling with a smaller amount of labeled data.

A key consideration for active learning is the sampling strategy used by the sheet selector for the selection of sheets for human labeling. The desirable strategy should make effective usage of the labeled data to optimize sample efficiency in learning. Thus, we employ a sampling strategy to select the most uncertain sheets in each iteration for the human labeler. The inclusion of uncertain samples from current iteration into the training set is more likely to achieve higher gains in learning the detector for the next iteration. 

We propose six measures below for the evaluation of sheet uncertainty.

\begin{itemize}
\item \emph{Classification uncertainty score}: One minus the average classification probability returned by the softmax values for all detected table regions on the sheet.
\item \emph{Mismatch score of segmentation and detection masks}: One minus the IoU between the segmentation mask and detection mask. The detection mask is produced by setting the values of all cells inside the detected table region to 1 and 0 otherwise. 
\item  \emph{Table/Sheet-level Sparsity factor}: Table-level sparsity factor is given by the ratio of blank cells in the detected table region, while sheet-level sparsity is given by the lowest sparsity factor for 
	all tables detected on the sheet.
\item \emph{Overlapping region indicator}: 1 if there is overlapping between any two detected table regions and 0 otherwise.
\item \emph{Boundary mismatch indicator}: 1 if there is boundary mismatch and 0 otherwise. A mismatch is identified if any detected boundary is on a blank column or row.
\item \emph{Out-of-region coverage ratio}: The ratio of the number of non-blank cells outside the detected table regions to the total number of non-blank cells.
\end{itemize}

Note that the first two measures above are based on the detector output, and the remaining ones are based on domain-specific rules. The overall sheet uncertainty metric is then given by the $L_2$ norm of the 6-dimensional vector comprising the six measures above. A high score for the uncertainty metric indicates low confidence in the detector output result, and/or the output is incompatible with our domain knowledge. Thus, we can select the unlabeled sheets by thresholding on the overall uncertainty score. The active learning algorithm is described in Algorithm 1. The stop criteria are met when model accuracy exceeds a predefined threshold or does not change over a few consecutive iterations.

\begin{algorithm}[!ht]
\begin{algorithmic}
\STATE \textbf{Require}: \scalebox{0.9}{Spreadsheet set $\mathcal{X} = \{x\}$, domain-specific rules $\theta$} \\
\STATE \textbf{Ensure}: \scalebox{0.9}{Table detector $D(\mathcal{T}$)} \\
\STATE \hspace{\algorithmicindent} \scalebox{0.9}{1: Initialize labeled sheet set $\mathcal{T} = \{\}$, and detector $D(\mathcal{T})$} \\
\STATE \hspace{\algorithmicindent} \scalebox{0.9}{2: Build sheet selector $S$ based on $D(\mathcal{T})$ and $\theta$}\\
\STATE \hspace{\algorithmicindent} \scalebox{0.9}{3: \textbf{repeat} }\\
\STATE \hspace{\algorithmicindent} \scalebox{0.9}{4:\hspace{\algorithmicindent}$\mathcal{S}$ selects subset $\mathcal{X}_L = \{x_L\}$ from $\mathcal{X}-\mathcal{T}$} \\
\STATE \hspace{\algorithmicindent} \scalebox{0.9}{5:\hspace{\algorithmicindent}Human labels on $\{x_L\}$ with table regions $\{r_L\}$,  } \\
\STATE \hspace{\algorithmicindent} \hspace{\algorithmicindent} \scalebox{0.9}{\text{                      update } $\mathcal{T}  \leftarrow \mathcal{T} \cup\{(x_L, r_L)\}$}\\
\STATE \hspace{\algorithmicindent} \scalebox{0.9}{6:\hspace{\algorithmicindent}Train table detector $D'(\mathcal{T} )$ on $\mathcal{T} $, update detector} \\
\STATE \hspace{\algorithmicindent} \hspace{\algorithmicindent}\hspace{\algorithmicindent} \scalebox{0.9}{$D(\mathcal{T} ) \leftarrow D'(\mathcal{T} )$}
\STATE \hspace{\algorithmicindent} \scalebox{0.9}{7: \textbf{until} pre-defined stop criteria is met} \\
\STATE \hspace{\algorithmicindent} \scalebox{0.9}{8: \textbf{return} table detector $D(\mathcal{T})$} \\
\caption{Active learning for TableSense}
\label{alg_act}
\end{algorithmic}
\end{algorithm}

\section{Evaluation Results}
\label{evaluation}

\subsection{Implementation and Experiment Setup} \label{implementationand}
We customized ResNets \cite{he2016deep} as the backbone for TableSense, and the pooling layers are removed. The BBR module is combined with the PBR module to achieve accuracy promotion. Both the BBR and PBR modules contain three convolutional layers yet have different receptive fields and prediction targets.

\begin{figure}[!t]
\centering
\includegraphics[width=3in]{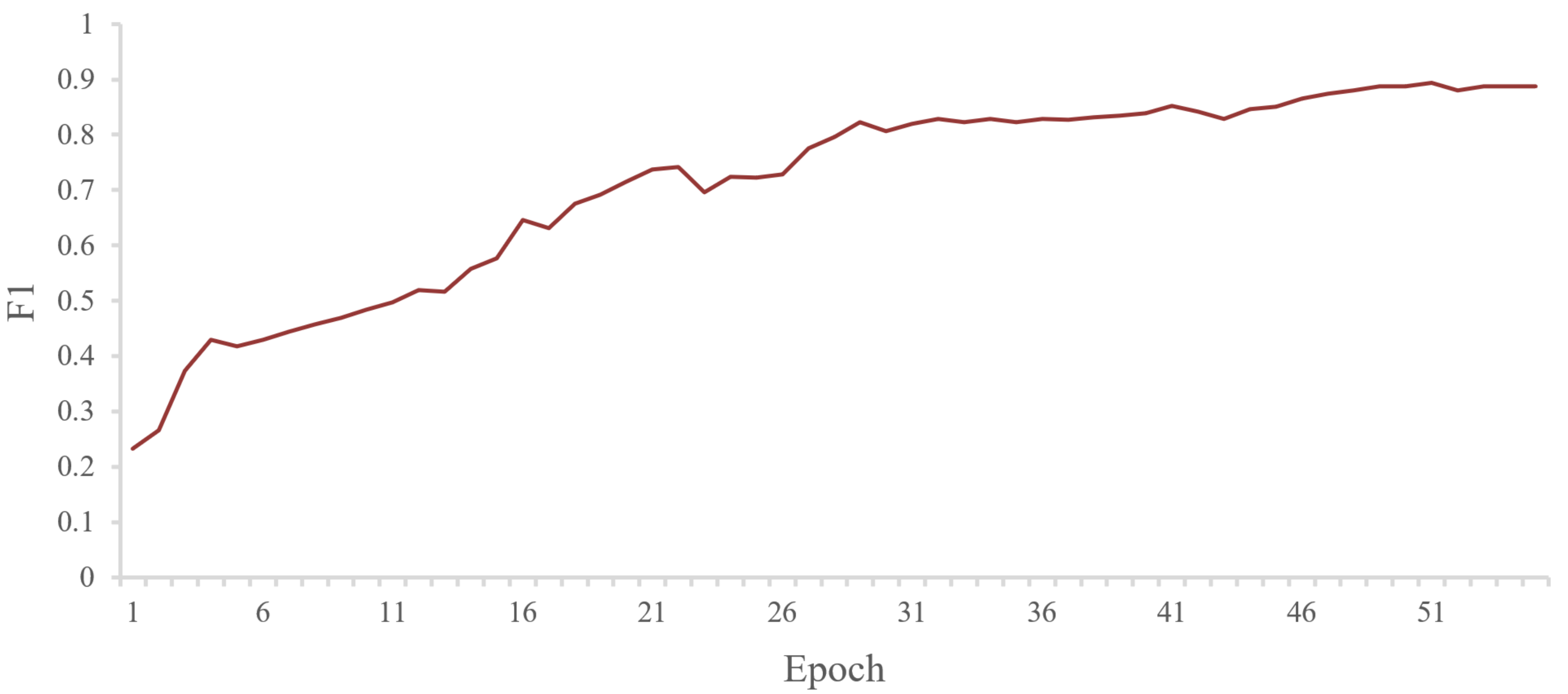}
\caption{Accuracy curve on test set WebSheet400.} \label{fig4}
\end{figure}

We use sheets in \textbf{WebSheet10K} for training and sheets in \textbf{WebSheet400} for testing. To parse Excel files and extract features, we use the ClosedXML\footnote{ https://github.com/ClosedXML/ClosedXML}  library. Since spreadsheets have various sizes, the mini-batch size for training is set to 1. Due to the large variations in sheet size, the span of RPN anchors and the span of aspect ratios in our model range from 8 to 4,096 and 1/256 to 256 incrementing by factor of 2 respectively. As a result, our model can detect small tables with only 12 cells up to large tables with over 100,000 cells. For the region proposal module, the proposed region number is set to 2,000, and the top 2,000 RoIs are further classified and refined in the detection branch. The weight decay is set to 0.0001 for regularization. The parameter $k$ for the PBR module is set to 7. The rescaled output size of RoIAlign is 14$\times$14. Our experiments are implemented on Nvidia V100 GPUs with TensorFlow \cite{abadi2016tensorflow}. Figure \ref{fig4} shows the accuracy curve on test set during training.
It can be clearly seen that the F1 scores for detection keep improving over the epochs, indicating the effectiveness of TableSense training for table detection. 
The trained model takes an average of 72ms for testing each sheet. 

\subsection{Effectiveness of Featurization}

We design our first comparison experiment to evaluate the effectiveness of different feature sets. The same training and testing flow of the TableSense model are repeated three times, for the binary valued features, features for value strings, and the full feature set in Table \ref{tab2} respectively. The evaluation results are shown in Table \ref{tab3}.

From the comparison results, we can see that proper cell feature representation plays a vital role in the effectiveness of the learning framework. To fully leverage the power of TableSense, cell features need to be properly designed to encode available information from diverse sources. Using a single feature set may lead to degenerate performance. Our feature set is highly effective for spreadsheet table detection.

\begin{table}

\caption{Precision and recall on WebSheet400 for TableSense based on different feature sets.}\label{tab3}
\scalebox{0.85}{
\begin{tabular}{l | c | c| c|c }
\whline

\% &  \multicolumn{2}{c|}{EoB-0} &\multicolumn{2}{c}{EoB-2}\\
\hline
Feature & Recall & Precision &  Recall & Precision \\
\hline
Binary value feature &  58.7  & 55.4 & 69.4 & 64.1 \\
Value string feature set  &  69.5 &70.2& 80.6 & 77.3\\
Full feature set & \textbf{80.8} &	\textbf{78.1}&	\textbf{91.3}&	\textbf{86.5}\\
\whline
\end{tabular}}
\end{table}

\subsection{Effectiveness of TableSense Architecture} \label{ex1}

We invest both the table detection functionality in existing commodity spreadsheet tools as well as the state-of-the-art method for object detection in computer vision. So we compare our method with these algorithms below.

\begin{itemize}
\item\textbf{Region-growth}, the existing table detection feature in Excel is based on region-growth with stride 1. Once a user chooses a cell and press Ctrl+*, Excel will automatically find the table range by expanding the area to include non-blank neighboring cells based on 8-connectivity.
\item \textbf{Region-growth + SVM}, we combine multi-stride region-growth and SVM, where an SVM model is trained to predict whether the result of region-growth corresponds to a true table region.
\item \textbf{Mask R-CNN}, the state-of-the-art deep learning based object detector.
\item  \textbf{YOLO-v3}, the state-of-the art one-stage object detector with high speed/accuracy trade-off. 
\item  \textbf{Faster R-CNN}, one of the state-of-the-art object detectors based on two-stage, proposal-driven mechanism.
\end{itemize}

\begin{table}
\caption{Spreadsheet table detection on WebSheet400.}\label{tab4}
\scalebox{0.85}{
\begin{tabular}{l | c | c| c|c }
\whline

\% & \multicolumn{2}{c|}{EoB-0} &\multicolumn{2}{c}{EoB-2}\\
\hline
Model & Recall & Precision & Recall & Precision \\
\hline
Region-growth & 43.4 &39.8 & 58.5 &52.2 \\
Region-growth + SVM & 44.3 & 54.3 & 66.5 & 56.7 \\
YOLO-v3 & 42.8 & 37.9 & 63.4 & 58.2 \\
Faster R-CNN & 46.0 & 41.4 & 64.7 & 65.3 \\
Mask R-CNN &48.4 &40.2 &75.5& 64.2\\
TableSense &\textbf{80.8} &\textbf{78.1} &\textbf{91.3} &\textbf{86.5}\\
\whline
\end{tabular}}
\end{table}

To ensure unbiased comparisons, all these methods use the same featurization scheme introduced in our approach.
The comparison results are reported in Table \ref{tab4}. The results show that Mask R-CNN, the state-of-the-art two-stage model, outperforms YOLO-v3, which is the state-of-the-art one-stage model for detection. Both of them are much better than region-growth in EoB-2 metric. However, with the stricter EoB-0 metric, even region-growth + SVM achieves a better precision than Mask R-CNN. This reflects the limitation of  straightforward applications of the state-of-the-art CNN-based approaches for precise boundary detection.

On the other hand, TableSense achieves a significant margin of improvement over all baselines. Compared to Mask R-CNN and other approaches, TableSense brings in even larger accuracy gain with EoB-0 than EoB-2. In addition, we also train TableSense without PBR for ablation study, but the EoB-2 recall drops from 91.3\% to 78.4\%, and precision drops from 86.5\% to 68.1\%. Thus, our proposed TableSense is highly effective regarding precise bounding box detection.

\subsection{Effectiveness of Uncertainty Metric for Active Learning}

To evaluate the effectiveness of the uncertainty metric for active learning, we simulate the sample selection process over all test sheets in WebSheet400 and investigate the uncertainty metrics for well detected sheets and sheets with errors. A sheet is well detected if and only if all tables on the sheet are correctly predicted. If active learning is effective, well detected sheets should have low uncertainty scores so that they are unlikely to be selected, and the case for sheets with errors should be opposite. Hence we define the accuracy metric for sample selection as the ratio of high uncertainty sheets for sheets with errors and the ratio of low uncertainty sheets for well detected sheets.

We have built the WebSheet10k training set in 7 iterations in total. In this experiment we evaluate the sample selection performance using the model after the 3rd iteration and the model after the last iteration. The results in Table \ref{tab5} show that both models achieve high selection accuracies for well detected sheets and sheets with errors. The performance of the model after 7th iteration is further improved over the model after 3rd iteration, as indicated by the accuracy values in Table \ref{tab5}. As a result, sheets with errors are much more likely to be selected for further labeling than well detected ones due to the contrast in uncertainty metrics. This clearly demonstrates the effectiveness of the proposed uncertainty metric for active learning. 
\begin{table}

\caption{Uncertainty selection on WebSheet400 after the 3rd and 7th iterations.}\label{tab5}
\scalebox{0.76}{
\begin{tabular}{l| l l |l l}
\whline
\multirow{2}{*}{\#Spreadsheets} &\multicolumn{2}{c|}{Model after 3rd iteration}	&\multicolumn{2}{c}{Model after 7th iteration}\\
\cline{2-5}
 &With error & Correct & With error & Correct\\
\hline
High uncertainty&	\textbf{154}	&28	&\textbf{83}&	42\\
Low uncertainty	&34	&\textbf{184}	&13&\textbf{262}\\
Total&	\textbf{188}&	\textbf{212}	&\textbf{96}&\textbf{304}\\
Selection accuracy	&\textbf{81.9\%}	&\textbf{86.7\%}&	\textbf{86.4\%}&	\textbf{86.1\%}\\
\whline
\end{tabular}}
\end{table}

\section{Related Work}
\textbf{Table Detection.} Considerable research has been done to extract tables from HTML \cite{wang2012understanding,zhai2005web,wang2002machine}, document images \cite{gatos2005automatic,zuyev1997table,hu1999medium,liu2008identifying,shafait2010table} and PDFs \cite{fang2012table,liu2007tableseer}. They all focus on the issue of extracting tables embedded in the ambient text and are well separated from the surroundings using techniques of image processing and meta data analysis. However, table detection in spreadsheets has largely been overlooked. \cite{doush2010detecting} adopts a rule-based approach, but there is still a big gap between table detection and structure analysis due to the great flexibility in crafting spreadsheet with diverse table structures and various tables layouts, so we devise TableSense to address these challenges. 

\textbf{R-CNN}. The Region-based CNN (R-CNN) \cite{girshick2014rich} is proposed to detect bounding boxes in images by proposing candidate regions \cite{uijlings2013selective,hosang2016makes} and evaluating convolutional networks \cite{lecun1989backpropagation,krizhevsky2012imagenet} for the RoIs. Fast R-CNN \cite{girshick2015fast} uses a RoIPool module to extend R-CNN by attending to RoIs on feature maps with improved detection performance. Faster R-CNN \cite{ren2015faster} extends Fast R-CNN by learning a Region Proposal Network (RPN) for the generation of candidate regions for detection. Mask R-CNN advances the stream by replacing the RoIPool with RoIAlign to preserve the explicit per-pixel spatial correspondence and adding a module for object mask prediction, achieving top results in tasks involving instance segmentation and object detection \cite{he2017mask}.

\section{Conclusion and Future Work}
In this paper, we propose the TableSense suite to address the challenges in spreadsheet table detection. TableSense is a unified, end-to-end framework customized from CNN with several key enhancements. First, we propose a featurization scheme to encode cell features. Second, we devise a PBR module to predict precise bounding boxes and incorporate it. Third, we use active learning to effectively select low confidence sheets for human labeling in building up the training dataset. In the future, we will leverage the TableSense technique for automated table structure analysis and make a further step in spreadsheet intelligence.

\bibliographystyle{aaai}
\fontsize{9.1pt}{10.1pt} \selectfont 
\bibliography{Reference_TableSense}

\end{document}